\documentclass[]{emulateapj}
\shorttitle{Breathing in Low Mass Galaxies}
\shortauthors{Stinson et al.}

\begin{document}

\title{Breathing in Low Mass Galaxies: A Study of Episodic Star Formation}

\author{G.\,S. Stinson\altaffilmark{1,2},
        J.\,J. Dalcanton\altaffilmark{1},
        T. Quinn\altaffilmark{1},
        T. Kaufmann \altaffilmark{3},
        J. Wadsley\altaffilmark{4} 
}
\altaffiltext{1}{Astronomy Department, University of Washington, Box 351580, Seattle, WA, 98195-1580}
\altaffiltext{2}{e-mail address: stinson@astro.washington.edu }
\altaffiltext{3}{Department of Physics and Astronomy, University of California, Irvine, Irvine, CA, 92697}
\altaffiltext{4}{Department of Physics and Astronomy, McMaster University, Hamilton, Ontario, L88 4M1, Canada}


\begin{abstract} 
We simulate the collapse of isolated dwarf galaxies using SPH + N-Body 
simulations including a physically motivated description of the effects of 
supernova feedback.  As the gas collapses and stars form, the supernova feedback disrupts enough gas to temporarily quench star formation.  The gas flows outward into a hot halo, where it cools until star formation can continue once more and the cycle repeats.  The star formation histories of isolated Local Group dwarf galaxies exhibit similar episodic bursts of star formation.  We examine the mass dependence of the stellar velocity dispersions and find that they are no less than half the velocity of the halos measured at the virial radius.

\end{abstract}


\keywords{galaxies: evolution --- galaxies: formation --- galaxies: dwarf --- methods: N-Body simulations}

\section{Introduction}
\label{intro}

Dwarf galaxies have shallow potential wells.  They are therefore subject to many baryonic processes that are negligible in high mass galaxies.  For example, dwarf galaxies are prone to gas loss through supernova feedback
and/or stripping.  They also are unable to efficiently accrete gas
whose temperature is higher than the virial temperature of the galaxy.
As a result, the baryonic content of dwarf galaxies is expected to be
lower than the cosmic mean, and should have significant scatter.

Given their susceptibility to galactic ``weather'', the cool baryonic
content of dwarf galaxies should fluctuate, not just from galaxy to
galaxy, but within an individual galaxy as a function of time.  The expected temporal
variations in the amount of cool gas should be accompanied by variations in the star formation rate (SFR).  These variations in the SFR can be seen within the Local Group, where dwarf galaxies are sufficiently well resolved
that their past SFRs can be directly inferred from
their color-magnitude diagrams.  

As expected, past star formation in
dwarf galaxies appears to be complex, with many systems
exhibiting multiple episodes of star formation
\citep[e.g.][]{Smecker-Hane96,DohmPalmer02, Rizzi04, Skillman05,
McConnachie06}.  The temporal separation of bursts of star formation
ranges from tens or hundreds of megayears in dwarf irregulars like
Sextans A, GR8 \& Phoenix \citep{DohmPalmer98, DohmPalmer02, Dolphin03,Young07}, to
gigayears in dwarf spheroidals like Carina \citep{Hurley-Keller98}.
In some systems these fluctuations have large amplitudes (e.g.\
Carina), and in others, the amplitudes are of the order of 50\% of the
mean star formation rate.  However, the true instantaneous fluctuation
amplitudes are possibly larger if the fluctuation is shorter than the
width of the time bins used for analysis.  The temporal resolution of
star formation history recovery methods worsens with increasing lookback time,
so it is possible that short timescale fluctuations are present at
early times as well.

These fluctuating star formation rates have
long been a puzzle.  Many of the galaxies observed to have episodic
star formation also have large intermediate age populations, showing
that they initially formed stars, then lay largely dormant, before
reviving several gigayears later \citep[e.g.\ Carina, DDO210, LeoA,
LeoI, IC1613;][]{Smecker-Hane96, Hurley-Keller98, McConnachie06,
Tolstoy98, Gallart99, Hernandez00, Skillman03}.  Others have no sign
of current star formation, but were actively forming stars within the
past 100 Myrs \citep[e.g.\ DDO210, DDO6, Fornax;][]{McConnachie06,
Weisz07}.

How is it possible that star formation in dwarf galaxies can stop and
start over a range of timescales?  Some of these systems are
relatively isolated (DDO210, LeoII, IC1613), ruling against interactions
as a driving force.  Some authors have speculated that reionization
may have held off significant SF until intermediate ages
\citep[e.g.][]{Bullock00,Skillman03,Gnedin06,Read06}, but such a mechanism
makes it hard to explain either the significant populations of truly
old stars \citep{Stetson98,Buonanno98,Grebel03} or the shorter timescale fluctuations
seen at the present day.  However, there is growing evidence from dwarf
spheroidal age-metallicity relations that gas accretion and outflows
may have played a role, leading to G- and K-dwarf problems comparable
to those seen in the Milky Way \citep{Koch06a, Koch06b}.

In this paper, we explore the role that self-regulation of feedback can play in
driving the observed fluctuations in the star formation rate, particularly on short timescales.  To do so,
we use high resolution numerical simulations of isolated galaxies,
including self-consistent star formation and SN feedback recipes
calibrated to match Milky Way sized galaxies \citep{Stinson06}.  We
maximize resolution and minimize the effect of interactions by
studying the evolution of uniform, collapsing spheres.  These
simulations are fully self-consistent, and thus differ from previous
efforts to model supernovae explosions in dwarf galaxies, most of which involve implanting
an energetic source into the galaxy and tracking the resulting winds \citep{DS86,
Burkert97,MLF99,Scan00, Scan01, Recchi01, Efstathiou00,Fragile03,
Fujita03, Pelupessy04,Hensler04,Marcolini06}.

Although examining galaxies outside of the full cosmological context sacrifices some details of the mass accretion history, it also allows our models to have sufficiently high resolution to capture the more relevant physics of the star formation and feedback processes.  The resulting smoothly collapsing halos are sufficiently simple so that we can isolate features of the galaxies' behavior that are due solely to the details of star formation and feedback.  

With these simplified but self-consistent models, we show that star
formation in low mass galaxies can undergo a natural ``breathing''
mode, where episodes of star formation trigger gas heating that
temporarily drives gas out of the cool phase and into a hot halo.
Subsequent gas cooling and gas accretion then allows gas to settle
back into the halo, and star formation begins again.  This
episodic mode of star formation is superimposed on the net infall of
gas into the galaxy, leading to galaxies with both episodic star
formation and significant intermediate age populations, in line with
observations.

In \S \ref{sec:methods}, we describe the smoothed particle hydrodynamics (SPH)
code, galaxy models, star formation, and supernova recipes used for the
simulations.  \S \ref{sec:results} shows the sequence of events that lead to the episodic
star formation we see in our simulations.  \S \ref{sec:compare}
compares the star formation in the simulations with that observed in Local Group dwarfs.




%
%

\section{Methods}
\label{sec:methods}

To simulate the star formation history in dwarf galaxies, we created five example galaxies in hydrostatic equilibrium with \citet{NFW} (hereafter NFW) density profiles.  They vary in total mass from $10^9 {\rm M}_\odot$ to $10^{10} {\rm M}_\odot$ and have velocities at the virial radius ($v_{200}$) of 10, 15, 20, 25, and 30 km s$^{-1}$ comparable to Local Group Dwarfs.  We do not consider halos smaller than $10^{9} {\rm M}_\odot$, which do not have enough mass to retain gas reionized by the UV radiation from the first stars and AGN.  

Although examining galaxies outside of the full cosmological context sacrifices some details of the mass accretion history, it also allows our models to have sufficiently high resolution to capture the more relevant physics of the star formation and feedback processes.  The resulting smoothly collapsing halos are sufficiently simple so that we can isolate features of the galaxies' behavior that are due solely to the details of star formation and feedback.  

\subsection{Models}
\label{sec:models}
We created combined gas and dark matter virialized halos using the specifications of \cite{Kaufmann07}.  The halos start with $10^5$ gas particles and $10^5$ dark matter particles, a resolution at which numerical losses of angular momentum become small in \citet{Kaufmann07}.  Equilibrium NFW dark matter halos with a concentration c = 8 were created following \citet{Stelios04}, where $c = r_{vir}/r_s$, $r_{vir}$ is the virial radius and $r_s$ is the scale radius.  The dark matter extends beyond the virial radius to keep the halos in equilibrium.  Dark matter inside the virial radius is at slightly higher resolution than outside such that 90\% of the dark matter particles are inside the virial radius, but only 72\% of the mass is.  

Gas was added to the halo following the same NFW density profile as the dark matter, but comprising only 10\% of the total halo mass.  Each gas particle was assigned an initial temperature to keep it in hydrostatic equilibrium with the halo before the gas begins radiatively cooling. All gas particles started with equal mass.  The force resolution is based on a fixed fraction of the virial radius ($r_{vir}$), so that the softening is set to $r_{vir}/2000$.  

Dark matter particles are left with their random equilibrium velocities while the gas is spun with a uniform circular velocity so that $\lambda=\frac{j_{gas}\left|E\right|^{\frac{1}{2}}}{GM^{\frac{3}{2}}}$ = 0.039, where $j_{gas}$ is the average specific angular momentum of the gas, $E$ and $M$are the total energy and mass of the halo.  No cosmological Hubble flow is included in the initial velocities.  

\subsection{Numerics}
The simulations in this paper were run using the parallel SPH code GASOLINE \citep{Gasoline}.  GASOLINE solves the equations of hydrodynamics, and includes the radiative cooling described in \citet{Katz96}.  The cooling assumes ionisation equilibrium, an ideal gas with primordial composition, and solves for the abundances of each ion species.  The scheme uses the collisional ionization rates reported in \citet{Abel97}, the radiative recombination rates from \citet{Black81} and \citet{Verner96}, bremsstrahlung, and line cooling from \citet{Cen92}. The energy integration uses a semi-implicit stiff integrator independently for each particle with the compressive heating and density
(i.e. terms dependent on other particles) assumed to be constant over the
timestep.  

Stars are formed and feedback is calculated every 1 Myr.  The star formation and feedback recipes are the ``blastwave model" described in detail in \citet{Stinson06}, and they are summarized as follows.  Gas particles must be dense ($n_{\rm min}=0.1 cm^{-3}$) and cool ($T_{\rm max}$ = 15,000 K) to form stars.  A subset of the particles that pass these criteria are randomly selected to form stars based on the commonly used star formation equation, 
\begin{equation}
\frac{dM_{\star}}{dt} = c^{\star} \frac{M_{gas}}{t_{dyn}}
\end{equation}
where $M_{\star}$ is mass of stars created, $c^{\star}$ is a constant star formation efficiency factor, $M_{gas}$ is the mass of gas creating the star and $t_{dyn}$ is the gas dynamical time.  This equation is integrated on 1 Myr star formation time scales in all of the simulations presented here. The constant parameter, $c^{\star}$ is tuned to 0.05 to match the \citet{Kenn98} Schmidt Law in the simulated Isolated Model Milky Way used in \citet{Stinson06}.

At the resolution of these simulations, star particles represent a large group of stars (300 M$_\odot$).  Thus, each particle has its stars partitioned into mass bins based on the initial mass function presented in \citet{Kroupa93}.  These masses are correlated to stellar lifetimes as described in \citet{Raiteri96}.  Stars larger than 8 $M_\odot$ explode as supernovae during the timestep that overlaps their stellar lifetime after their birth time.  Similarly, Type Ia supernovae are included for stars smaller than 8 $M_\odot$.  The explosion of all supernovae is treated using the analytic model for blastwaves presented in \citet{MO77} as described in detail in \citet{Stinson06}.  While the blast radius is calculated using the full energy output of the supernova, only half of that energy is transferred to the surrounding ISM, $E_{SN}=5\times10^{50}$.  The other half of the supernova energy is radiated away.

\section{Results}
\label{sec:results}
We ran simulations of the five isolated, low mass, virialized galaxy halos for 10 Gyr (approximately 30 dynamical times).  In the lowest mass galaxies, we observed that the star formation varies episodically.  Other groups have noted similar behavior.  \citet{Dong03} produced episodic star formation in one dimensional simulations.  \citet{Pelupessy04} starts SPH simulations at an advanced stage where there is a significant population of stars, and all the gas is in the disk.  Using SN feedback similar to what is employed here, \citet{Pelupessy04} still note episodic star formation in model galaxies with similar mass to our 15 km s$^{-1}$ halo.  

The observed fluctuations are driven by supernova feedback.  As gas accretes and stars form, the resulting feedback is sufficient to resist the collapse of gas and drive it outwards.  The subsequent drop in gas density shuts off the burst of star formation.  However, supernova remnants cool and gas can once again accrete and another wave of stars can form.  We now discuss this behavior in more detail.

\begin{figure}
\resizebox{9cm}{!}{\includegraphics{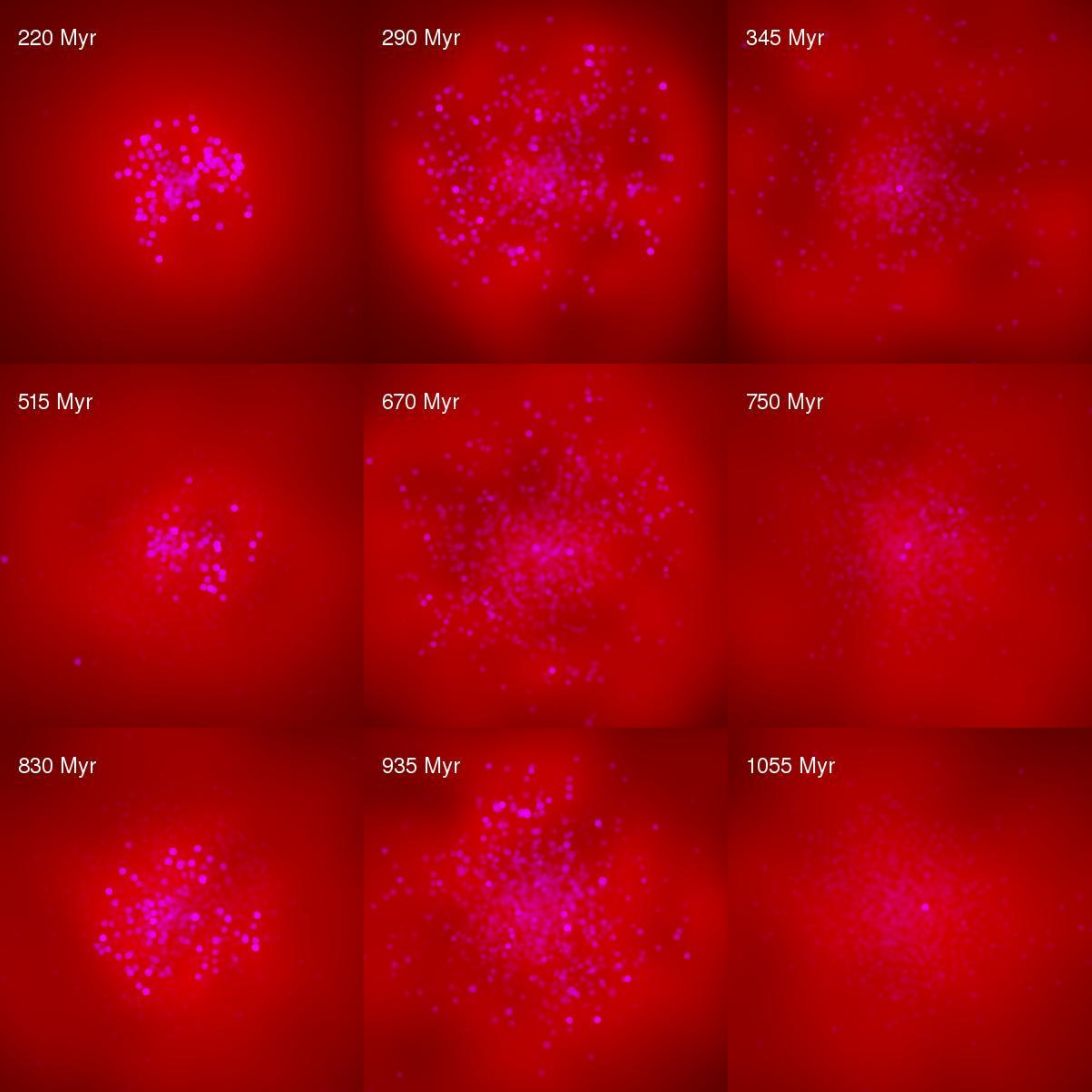}}
 \caption{ The sequence of events that produces episodic star formation in the 15 km s$^{-1}$ run.  Purple dots represent the stars.  A star burst in a dense gaseous region (colored bright red) causes the outflow of gas that prevents more stars from forming for a period of time until the gas is able to cool once more.  The cycle repeats itself as there is a new star burst that drives gas and ends star formation.  }
\label{fig:pics} 
\end{figure}

Figure \ref{fig:pics} shows the sequence of events illustrating the episodic behavior in a single galaxy.  First, the infall of cooling gas triggers a central star burst (shown in the left panels).  The subsequent supernova feedback contains enough energy to eject gas from the shallow potential well of the small galaxy halo (center panels).  As the gas expands adiabatically, the density of the gas in the center drops below the star formation threshold (n = 0.1 $cm^{-3}$) (right panels).  The cessation of star formation activity then allows gas to cool, contract and again become dense enough to form stars.  The process then repeats in another cycle of star formation, expansion and cooling.  In this way, an episodic star formation history is created in an isolated galaxy without triggering by external forces.  Timescales and amplitudes for this episodic behavior are given in Table \ref{tab:data}, as a function of galaxy mass.

\begin{figure}
\resizebox{9cm}{!}{\includegraphics{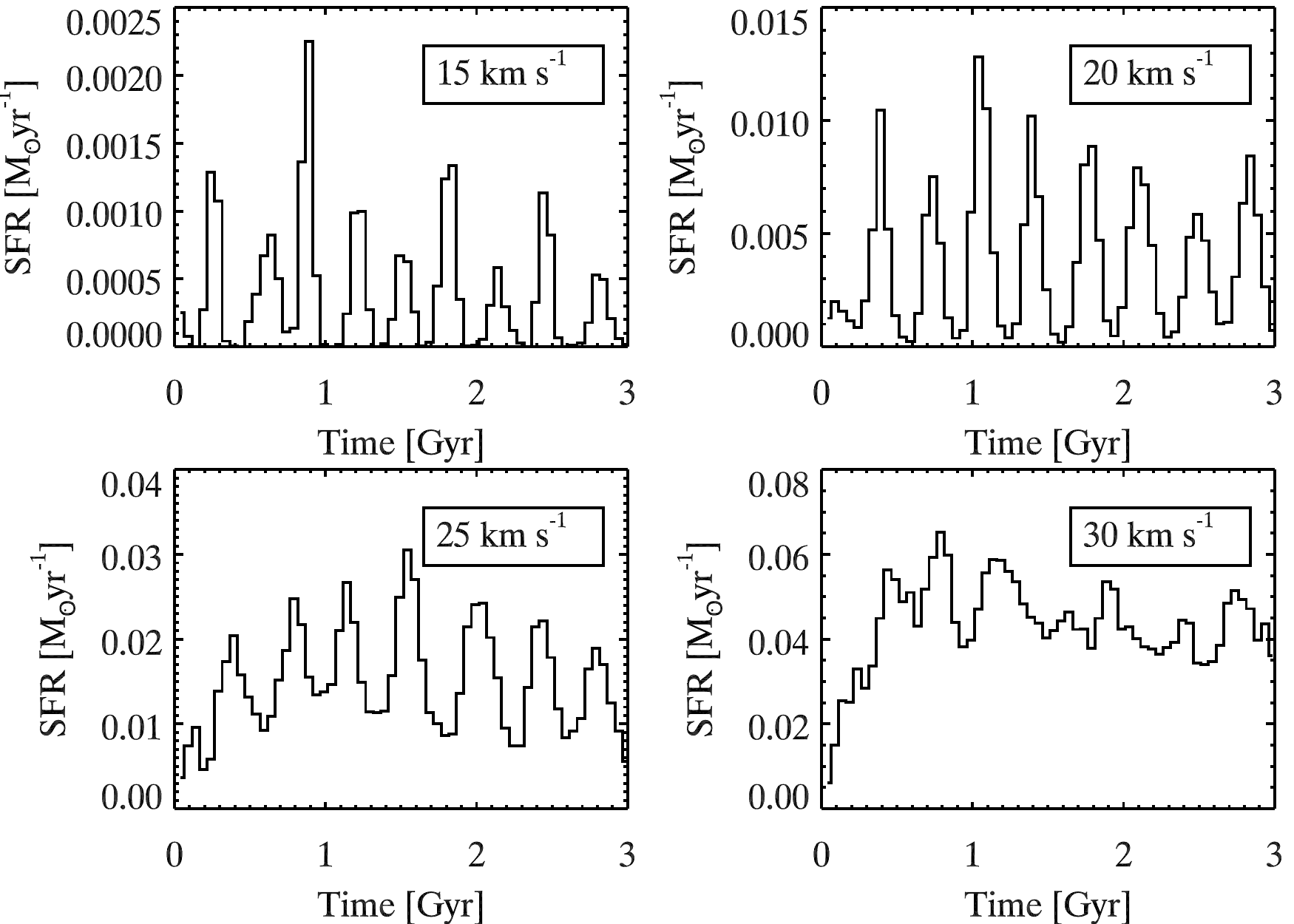}}
\caption{The star formation history of four dwarf galaxies. The star formation histories vary as the mass of the galaxy increases, so that there are fewer, more widely separated fluctuations in more massive galaxy potentials. }
\label{fig:sfr}
\end{figure}

Figure \ref{fig:sfr} shows the star formation histories of the four most massive halos.  The lowest mass halo is excluded because its gas cannot cool efficiently and forms few stars.  The star formation clearly fluctuates in the 15, 20, and the 25 km s$^{-1}$ halos.  In these isolated, idealized galaxies, the fluctuations are strictly periodic, though they would be unlikely to be so in a full cosmological context.  The periods of their oscillation are 292, 342, and 387 Myr, respectively.  These times are characteristic of the dynamical time of the central region of constant density gas that forms after the starburst (250 Myr for the 15 km s$^{-1}$ halo).  The dense core of the galaxies can cool in less than 100 Myr, which is much shorter than the dynamical time.  Thus cooling does not significantly impede the collapse of the galaxy and formation of stars and thus the period is instead dominated by the dynamical time for the gas to contract. 

For larger galaxies, star formation never completely ceases due to the reduced effectiveness of feedback in deeper potential wells.  Some episodic behavior is seen, superimposed above a constant level and with a slightly longer period.  This constant star formation exerts an outward pressure on the gas, slowing the rate of infall and cooling of the gas, and increasing the separation between bursts.

\begin{figure}
\resizebox{9cm}{!}{\includegraphics{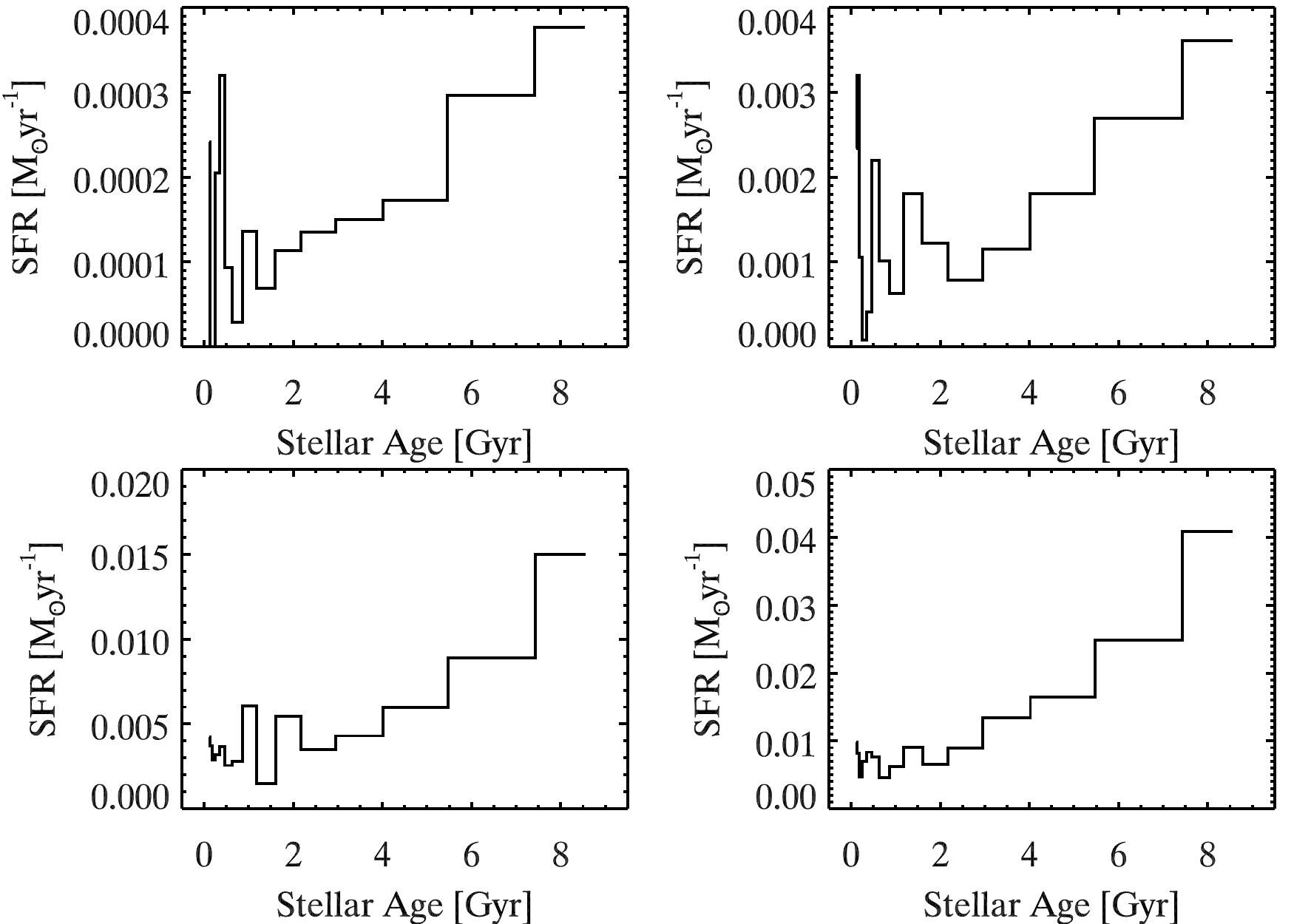}}
\caption{The star formation rate as a function of stellar age where the age of the stars has been averaged over log bins as a rough comparison to observations.  The newest stars on the left have the smallest bins to reflect observations where recent stellar ages are more easily resolved than old ages.}
\label{fig:ages}
\end{figure}

\begin{deluxetable}{c c c c c c c}
\tablecaption{Simulation data}
\tablewidth{0pt}
\tablehead{
\colhead{M$_{tot}$}&\colhead{v$_{200}$}&\colhead{$\sigma_v$}&\colhead{M$_\star$}&\colhead{P$_{SFR}$}&\colhead{SFR}&\colhead{b} \\
\colhead{(M$_\odot$)} & \colhead{(km} & \colhead{(km}&\colhead{(M$_\odot$)}&\colhead{(Myr)}&\colhead{Amplitude}&\colhead{} \\
\colhead{} & \colhead{s$^{-1}$)} & \colhead{s$^{-1}$)}&\colhead{}&\colhead{}&\colhead{}&\colhead{} 
}
\startdata
$3.18 \times 10^8$&10&4.5&$1.23\times10^4$&292&4.17&0\\
$10^9$&15&7.4&$2.34\times10^6$&292&2.98&0.434\\
$2.5 \times 10^9$&20&11.4&$2.15\times10^7$&342&2.25&0.526\\
$5 \times 10^9$&25&15.1&$7.86\times10^7$&387&0.73&0.549\\
$8.6 \times 10^9$&30&20.1&$2.20\times10^8$&-&0.42&0.539\\
\enddata
\tablenotetext{1}{  M$_{tot}$ is the virial mass of the halo including dark and baryonic matter.}
\tablenotetext{2}{  v$_{200}$ is the circular velocity at the virial radius to which the total mass corresponds.  }
\tablenotetext{3}{$\sigma_v$ is the rms of the velocity distribution.  }
\tablenotetext{4}{M$_\star$ is the stellar mass that accumulates over 10 Gyr.  }
\tablenotetext{5}{P$_{SFR}$ is the time period corresponding to the strongest mode in a Fourier decomposition of the star formation history.  }
\tablenotetext{6}{SFR Amplitude is maximum to minimum peak difference of the star formation history divided by the mean star formation rate. }
\tablenotetext{7}{b represents the median Scalo b (SFR/$<$SFR$_{\rm past}>$)parameter value.}
\label{tab:data}
\end{deluxetable}

In addition to the slight change in period, the amplitude of SFR variations decreases with mass.  At higher masses, the supernova energy is a smaller fraction of the galaxy's gravitational potential.  




\section{Comparison with Observations}
\label{sec:compare}
To facilitate comparison with observations, we have translated the SFHs shown in Figure \ref{fig:sfr} into the form most commonly derived from observed color magnitude diagrams.  Star formation history recovery methods typically use bins that increase logarithmically with time as shown in Figure \ref{fig:ages}.

Overall, our simulations look remarkably similar to actual observations.  For instance, compare Figures 6 and 7 of \citet{Young07} to our 15 km s$^{-1}$ star formation histories in Figures \ref{fig:sfr} and \ref{fig:ages}.  Like real low mass galaxies, the model galaxies show significant star formation at the present day and have not suffered from overcooling.  Most importantly, however, they show significant variation in the SFR at the present day (similar to the simulations of \citet{Pelupessy04} and observations of Sextans A \citep{DohmPalmer98}, GR8 \citep{DohmPalmer02}, and Phoenix \citep{Young07}), in spite of the total absence of external triggers.

We have modeled the galaxies colors, luminosity and surface brightness using SUNRISE \citep{Jonsson06}.  SUNRISE is a radiative transfer program that ray traces stellar emission through a grid that scatters light from dust, based on the metallicity of gas present in the simulation.  It is the most realistic converter from simulation to observational data available today.  The absolute magnitudes of the model galaxies agree well with the real low mass galaxies (e.g. 15 km s$^{-1}$ halo has M$_V$=-10.7, comparable to Phoenix with M$_V$=-10.1 \citep{Mateo98}).  We have also examined how the color and magnitude varies during the burst cycle.  We find that the $g-r$ color (which should be insentitive to emission lines \citep{West05}, varies by $\sim 0.2$ magnitudes during the burst and quiescent phase.  The magnitude drops by only 0.5 magnitudes before another burst ignites, suggesting the galaxies should remain visible during the quiescent phase.

Where the models and real galaxies differ, however, is in the star formation rates at intermediate times.  The models all show star formation rates that decline to the present day on average.  However, real low mass galaxies tend to have nearly constant average star formation rates and sometimes have star formation histories that peak at intermediate age.  This discrepancy is not surprising given that our models have been set up without any ongoing source of gas accretion.  Galaxies in a full cosmological context will be capable of much more varied star formation histories.  Our simulations suggest that very long separations of star formation episodes (such as in Carina) are unlikely to be due to feedback.  Instead, the recurrence of star formation must be due to the physics left out of our simulations, namely interactions and ongoing gas accretion from the cosmic web.

\section{Conclusions} 
Our study of isolated, spherical, collapsing low mass galaxies shows that star formation rates do not remain constant in galaxies the size of Local Group dwarfs.  An effective supernova feedback mechanism is able to drive gas out of the galaxies and create periodic episodes of star formation.  In the simple simulations presented here, the star formation followed regular, periodic oscillations.  However, observations also show additional variability on longer time scales including inactivity for many Gyrs.  It is not hard to imagine that such fluctuations are possible in the cosmological context of the Universe where galaxies may accrete gas sporadically over a long period of time and ocasionally interact with other structures in the cosmic web.  


Future work will discuss the outflows from low mass galaxies and how star formation scales in similar galaxies with higher masses.


\acknowledgments

We would like to thank Chris Brook and Patrik Jonsson for helpful 
conversations during this project.  
GS and TQ were supported by NSF ITR grant PHY-0205413.  
JD was supported NSF CAREER AST-0238683.
The Condor Software Program (Condor) was
developed by the Condor Team at the Computer Sciences Department of
the University of Wisconsin-Madison. All rights, title, and interest
in Condor are owned by the Condor Team.  The galaxy simulations were
run on machines funded by the Student Technology Fee of the University
of Washington.  




\end{document}